\documentclass[fleqn,twoside]{article}
\usepackage{espcrc2}

\usepackage{graphicx}
\usepackage{epsfig}
\usepackage[figuresright]{rotating}

\newcommand{\AmS}{{\protect\the\textfont2
  A\kern-.1667em\lower.5ex\hbox{M}\kern-.125emS}}

\title{Conceptual design of a scalable multi-kton superconducting magnetized
liquid Argon TPC}

\author{A. Ereditato\address[INFN]{Istituto Nazionale di Fisica Nucleare, INFN Sezione di Napoli, Napoli, Italy}  and
        A. Rubbia\address[ETHZ]{Institut f\"{u}r Teilchenphysik, ETHZ, CH-8093 Z\"{u}rich,
Switzerland}}
       
\begin{document}

\begin{abstract}
We discuss the possibility of new generation neutrino and astroparticle physics experiments 
exploiting a superconducting magnetized liquid Argon Time Projection Chamber (LAr TPC).
The possibility to complement the features of the LAr TPC
with those provided by a magnetic field has been considered in
the past and has been shown to open new physics
opportunities, in particular in the context of a neutrino factory.
The experimental operation of a magnetized 10~lt LAr TPC prototype
has been recently demonstrated.
From basic proof of principle, the main challenge to be addressed is
the possibility to magnetize a very large volume of Argon, corresponding to 10~kton or more, 
for future neutrino physics applications. In this paper we present one such conceptual design.
\vspace{1pc}
\end{abstract}

\maketitle

\section{Introduction}

The liquid Argon Time Projection Chamber was
proposed~\cite{intro1} as a tool for uniform and high accuracy imaging of massive detector volumes. 
The chamber is based on
the fact that in highly pure Argon, ionization tracks can be drifted
over distances of the order of meters. 
Imaging is provided by wire planes at the end of the drift path, continuously recording the 
signals induced. $T_0$ is provided by the prompt scintillation light. 

\begin{figure*}[htb]
\centering
\epsfig{file=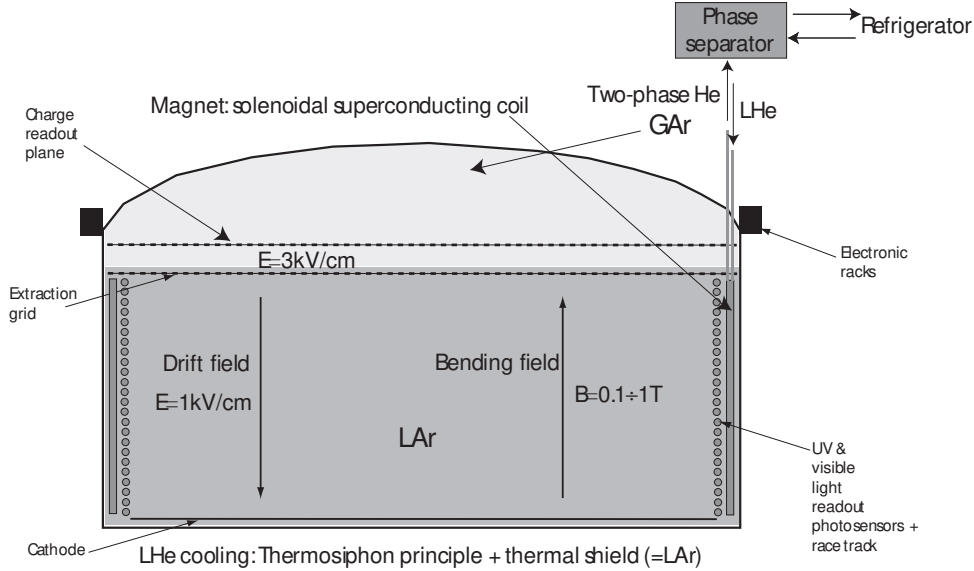,width=0.8\textwidth}
\vspace{-1cm}
\caption{\small Conceptual design of a superconducting
magnetized 100~kton liquid Argon detector.
The cryogenic tanker was developed
in collaboration with Technodyne International Limited.
This design can be scaled down to $e.g.$ 10~kton.}
\label{fig:glacier}
\end{figure*}

The main technological challenges are summarized
elsewhere~\cite{t600paper} and include techniques of Argon purification,  operation of wire 
chambers in cryogenic liquid and without charge amplification, 
low-noise analog electronics, continuous wave-form recording and digital signal processing.
The extensive ICARUS R\&D program dealt with
studies on small LAr volumes, LAr purification methods, readout schemes and electronics, 
as well as studies with several prototypes of increasing mass on purification technology,
collection and analysis of physics events, long duration tests and
readout~\cite{3tons,Cennini:ha,50lt}. 
The test of a 300 ton module 
carried out at surface~\cite{t600paper} has demonstrated that
the technique can be operated at large mass scale with a drift length of 1.5~m.
Data taking with cosmic-ray events assessed
the detector performance in a quantitative way (see references in~\cite{t600paper}),
in view of its installation at the LNGS.

\begin{table*}[htb]
\caption{Comparison of superconducting solenoidal magnets. ATLAS column corresponds
to the solenoid.}
\label{table:magpars}
\begin{tabular}{@{}|l|c|c|c|c|}
\hline
	& 10 kton LAr	&		100 kton LAr		&	ATLAS & CMS \\
Magnetic induction (T) & 0.1/0.4/1.0  & 0.1/0.4/1.0 &2.0&4.0 \\
Solenoid diameter (m) &30	&		70		&	2.4	&6\\
Solenoid length (m) &10		&	20	&		5.3&	12.5\\
Magnetic volume (m$^3$)&	7700	&		77000	& 21& 400\\
Stored magnetic energy (GJ)&	0.03/0.5/3 &	0.3/5/30	&0.04 & 2.7\\
Magnetomotive force (MAt)&	0.8/3.2/8 & 1.6/6.4/16	&9.3 &42\\
Radial magnetic pressure (kPa)&	4/64/400	& 4/64/400 & 1600&  6500\\
Coil current (kA) & 	\multicolumn{2}{c|}{30   (I/Ic=50\%)}		& 8 & 20\\
Total length conductor (km)&	2.5/10/25	& 12/57/117& 5.6 & 45\\
Conductor type	& \multicolumn{2}{c|}{NbTi/Cu normal superconductor, T=4.4K} & NbTi/Cu& NbTi/Cu\\
& \multicolumn{2}{c|}{or HTS superconductor ? (see text)} & & \\
\hline
\end{tabular}
\end{table*}

The possibility to complement the features of the LAr TPC
with those provided by a magnetic field has been considered in
the past~\cite{Bueno:2001jd,Rubbia:2001pk} and would open new
possibilities: (a) charge discrimination, (b)
momentum measurement of particles escaping the detector ($e.g.$ high energy muons), (c)
very precise kinematics, since the measurement precision is limited by multiple
scattering ($e.g.$ $\Delta p/p\approx$4\% for a track length of $L\!=\!12$\,m and a field  of
$B\!=\!1$\,T). 
The orientation of the magnetic field can be chosen such that the bending direction is
in the direction of the drift where the best spatial resolution is achieved. This is
possible since the Lorentz angle is small~\cite{Badertscher:2004py,Badertscher:2005te}. However, this is not
mandatory and the B--field could also be parallel to the drift field.

The presence of magnetic field is certainly
beneficial for the application in the context of the neutrino factory,
depending on the actual field strength~\cite{Rubbia:2001pk}: 
(1) a low field, $e.g.$ B=0.1~T, for the measurement of 
the muon charge (CP-violation); 
(2) a strong field, $e.g.$ B=1~T for the measurement of the muon/electron charges (T-violation).
For a practical application, however, the mass of the detector will have to be very large,
in the multikton range.
We present in this paper a design for a superconducting magnetized liquid Argon TPC,
based on our design for a large mass detector, described in the next section.

\section {A large mass liquid Argon TPC detector with charge imaging and light readout}
The success of the fully industrial construction of the ICARUS 
T600 module and its performance
in the surface test run has further motivated the modular
idea of ``cloning'' the T600 module
to reach the multi-kton scale.
However, modularity was not imposed by the
LAr TPC technique itself, but was an implementation choice motivated by the 
boundary conditions of the LNGS laboratory
and by the requirement to build the detector outside of the underground hall. 

On the other hand, we have already presented
our conceptual design for a monolithic 100~kton LAr TPC 
elsewhere~\cite{Rubbia:2004tz}. 
Such a scalable, single LAr tanker design is the most attractive solution from the point of view
of physics, detector construction, operation and cryogenics, and finally cost. 
Our studies show that the maximum size of the single unit is limited by the requirement
to locate the detector in an underground cavern, and not by the cryogenic tanker
itself. On the other hand, the design can also be scaled down to 10~kton, or even 1~kton.

For an updated status of this program, see~\cite{Ereditato:2005ru}.
 
\section {A superconducting magnetized large mass liquid Argon TPC detector}

We have recently successfully operated the first experimental prototype of a magnetized 
liquid Argon TPC~\cite{Badertscher:2004py,Badertscher:2005te}.
These encouraging results allow us to further consider a
large detector with magnetic field.
Beyond the basic proof of principle, the main challenge to be addressed is
the possibility to magnetize a very large mass of Argon, in a range of 10~kton or more, 
for future neutrino physics applications. 

The most practical design, which fits the concept described in the previous Section,
is that of a vertically standing solenoidal coil producing vertical field lines, parallel
to the drift direction.
Hence, {\bf we propose to magnetize the very large LAr volume by immersing a superconducting 
solenoid directly into the LAr tank} to create a magnetic field, parallel to the drift-field
(See Figure~\ref{fig:glacier}).

A magnetized LAr TPC was already addressed by Cline et al.~\cite{cline03}.
However, the presence of long wires disfavored the use of a magnetic field. 
In addition, the proposed warm coil would dissipate 17~MW at B=0.2~T (88 MW @ 1~T).
Such heat, even if affordable, would impose strong technical constraints
on the thermal insulation of the main tank. 
In contrast, superconductors produce no heat dissipation and the coil current 
flows even in absence of the power supply.

Can such large volume solenoids be built?
In Table~\ref{table:magpars} we summarize the relevant physical parameters 
for a 10 and 100~kton liquid Argon detectors for three different magnetic
field configurations (0.1, 0.4 and 1.0~T), compared
to existing solutions for LHC experiments.  Of course, a very large amount of energy will be stored
in the magnetic fields. However, owing B$^2$ dependence of the energy density,
the total amount of magnetic energy is comparable to that of
LHC experiments. For example, the magnetic volume of
a 10~kton liquid Argon detector is 7700~m$^3$, 
to be compared to 400~m$^3$ for CMS,
 $i.e.$ an increase by a factor $\simeq 20$. However, the CMS
field is 4~T, hence,  a 10~kton LAr detector with 
a field B$\approx 4/\sqrt{20}\approx 1~$T has
the same stored energy.
Since most issues related to large magnets 
(magnetic, mechanical, thermal) scale with the stored energy, we
readily conclude, that {\bf the magnetization
of very large liquid Argon volumes, although certainly very challenging,
is not unrealistic}! More figures of comparison
are shown in Table~\ref{table:magpars}.

In the case of quenching, 
the liquid Argon also plays the role of the thermal bath.
In the largest considered volume, $i.e.$ for a 100~kton detector, and
for a B=0.1T (resp. 1T), the stored energy in the B-field is 280 MJ (resp. 30 GJ). 
In case of full quenching of the coil, the LAr would absorb the dissipated heat 
which would produce a boil-off of 2 tons (resp. 200 tons) of LAr. 
This corresponds to 0.001\% (resp. 0.1\%) of the total LAr contained in the tank 
and hence supports the design.
			
\section {The use of High $T_C$ superconductors?}
A new era in superconductivity opened in 1986 when Bednorz and Mueller in Zurich discovered superconductivity at a temperature of approximately 30~K. In the following years, new materials were found and currently the world record is T$_c\approx$130~K.
HTS are fragile materials and are still at the forefront of material science research. 
However, they hold large potentials for future large scale applications.
For example, ribbons of BSCCO-2212 ($Bi_2-Sr_2-Ca_2-Cu_3-O_{10}$) with $T_c=$110 K are 
very promising and are regularly produced by $e.g.$ American 
Superconductor~\cite{amsuper}. 
Tapes of Bi2223 or YBCO are promising 2nd generation
HTS cables which will find many
applications in the next years. Unfortunately, HTS superconductors are much
more difficult to use than normal superconductor. In particular, their conductivity
depends strongly on temperature and on external perpendicular magnetic fields.
Nonetheless, magnets have been constructed with HTS.
The development of  Superconducting Magnetic Energy Storage (SMES)
systems will surely drive the technical developments of these HTS ribbons
in the years to come.
Today it is conceivable to estimate that very large coils from 
BSCCO-2212 could be operated at T=10~K,
with BSCCO-2223 @ T=20~K, and possibly YBCO @ T=50~K(?).
We have started an R\&D program to prove the conceptual feasibility of this 
idea~\cite{strauss}.

\section {Tentative yoke parameters}

In order to close the magnetic field lines, one can think of a simple
cylindrical and hollow iron ring to encompass the liquid Argon
volume. The thickness of the iron ring is determined by the magnetic
flux that needs to be absorbed. We summarize in Table~\ref{table:magyoke} the 
relevant dimensions.

In the case of SMES considered for underground storage of MJ energy,
systems without return yoke buried in tunnels in bedrock have
been considered and studied~\cite{eyssa}. If an underground location where stray
field can be tolerated is considered, then one could avoid using a yoke.

\begin{table}[tb]
\caption{Tentative parameters for return iron yoke.}
\label{table:magyoke}
\begin{tabular}{@{}lccc}
\hline
Cyl. Fe yoke &	10 kton LAr		&	100 kton LAr\\
\hline
Magn. ind. (T)&	0.1/0.4/1.0	&0.1/0.4/1.0\\
Magn. flux (W)&	70/280/710	&385/1540/3850\\
Field in Fe (T)&	1.8		&	1.8		\\
Thickness (m)&	0.4/1.6/3.7	&1/3.7/8.7\\
Height (m)&	10		&	20		\\
Iron Mass (kton)	&6.3/25/63 &34/137/342\\
\hline
\end{tabular}
\vspace{-1.0cm}
\end{table}

\section {What next? A 50~ton magnetized prototype?}
Before very large detectors could be considered, a basic proof of
the charge and momentum measurements via magnetic analysis
should be experimentally demonstrated. To this goal, we point
out that a $\approx$~50~ton detector in a magnetic field exposed
to (1) a charged particle beam ($e^\pm$, $\mu^\pm$, $h^\pm$,...)
(2) neutrino beam would be well adequate for such measurement
campaign.

\section {Conclusion}
In this paper, we have presented the conceptual design
of a new generation neutrino and astroparticle physics detector 
exploiting a superconducting magnetized liquid Argon Time Projection Chamber.
This detector would open unique new physics
opportunities, in particular in the context of a neutrino factory.
We have
started an R\&D program along these lines of thoughts. The operation
of a magnetized $\approx 50$~ton LAr setup would represent
a major important milestone in this context.

\end{document}